\documentclass[aps,prl,10pt,twocolumn,longbibliography,noeprint,floatfix,superscriptaddress]{revtex4-2}

\usepackage{amsmath, amssymb}
\usepackage{mathtools}
\usepackage{lipsum}
\usepackage{times}
\usepackage{graphicx}
\usepackage{wasysym}
\usepackage{bm}
\usepackage{hyperref}
\usepackage{xspace}
\usepackage{siunitx}
\usepackage[dvipsnames]{xcolor}
\usepackage{soul}
\usepackage{centernot}
\usepackage{braket}
\usepackage{microtype}
\usepackage{upgreek}

\definecolor{customcolor}{rgb}{0.261, 0.212, 0.658}

\definecolor{myColor2}{rgb}{0.02,0.12,0.3}

\definecolor{myColor}{rgb}{0.02,0.12,0.7}
\definecolor{myciteColor}{rgb}{0.39,0.7,0.89}
\hypersetup{colorlinks=true, linkcolor=myColor, filecolor=myColor, urlcolor=myColor, citecolor=myColor, urlcolor=myColor}
\DeclareSIUnit{\gauss}{\ensuremath{\mathrm{G}}}
\DeclareSIUnit{\bohrradius}{\ensuremath{a_0}}
\graphicspath{{./Figures/}}

\DeclareSIUnit{\nK}{\nano\kelvin}
\DeclareSIUnit{\um}{\micro\metre}
\DeclareSIUnit{\aB}{\emph{a}_0}
\DeclareSIUnit{\G}{G}

\newcommand{\appropto}{\mathrel{\vcenter{
  \offinterlineskip\halign{\hfil$##$\cr
    \propto\cr\noalign{\kern2pt}\sim\cr\noalign{\kern-2pt}}}}}

\def\be{\begin{equation}}
\def\ee{\end{equation}}

\makeatletter
\def\@fnsymbol#1{\ensuremath{\ifcase#1\or *\or \dagger\or \ddagger\or
   \mathsection\or \mathparagraph\or \|\or **\or \dagger\dagger
   \or \ddagger\ddagger \else\@ctrerr\fi}}
\makeatother

\newcommand{\um}{\upmu{\rm m}}
\newcommand{\nK}{\textrm{nK}}
\newcommand{\kB}{k_{\textrm{B}}}

\newcommand{\Tc}{T_{\textrm {c}}}

\newcommand{\potassium}{^{39}\textrm{K}}

\begin{document} 
 
\title{
{\color{black}
Suppression and enhancement of bosonic stimulation by atomic interactions}
}
\author{Konstantinos Konstantinou}\email{kk688@cam.ac.uk}
\affiliation{Cavendish Laboratory, University of Cambridge, J. J. Thomson Avenue, Cambridge CB3 0HE, United Kingdom}
\author{Yansheng Zhang}
\affiliation{Cavendish Laboratory, University of Cambridge, J. J. Thomson Avenue, Cambridge CB3 0HE, United Kingdom}
\author{Paul H.~C. Wong}
\affiliation{Cavendish Laboratory, University of Cambridge, J. J. Thomson Avenue, Cambridge CB3 0HE, United Kingdom}
\author{Feiyang Wang}
\affiliation{Cavendish Laboratory, University of Cambridge, J. J. Thomson Avenue, Cambridge CB3 0HE, United Kingdom}
\author{Yu-Kun Lu}
\affiliation{Research Laboratory of Electronics, MIT-Harvard Center for Ultracold Atoms, Department of Physics, Massachusetts Institute of Technology, Cambridge, 02139 Massachusetts, USA}
\author{\\ Nishant Dogra}
\affiliation{Cavendish Laboratory, University of Cambridge, J. J. Thomson Avenue, Cambridge CB3 0HE, United Kingdom}
\author{Christoph Eigen}
\affiliation{Cavendish Laboratory, University of Cambridge, J. J. Thomson Avenue, Cambridge CB3 0HE, United Kingdom}
\author{Tanish Satoor}
\affiliation{Cavendish Laboratory, University of Cambridge, J. J. Thomson Avenue, Cambridge CB3 0HE, United Kingdom}
\author{Wolfgang Ketterle}
\affiliation{Research Laboratory of Electronics, MIT-Harvard Center for Ultracold Atoms, Department of Physics, Massachusetts Institute of Technology, Cambridge, 02139 Massachusetts, USA}
\author{Zoran Hadzibabic}
\affiliation{Cavendish Laboratory, University of Cambridge, J. J. Thomson Avenue, Cambridge CB3 0HE, United Kingdom}

\begin{abstract}
The tendency of identical bosons to bunch, seen in the Hanbury Brown--Twiss effect and Bose--Einstein condensation, is a hallmark of quantum statistics. This bunching can enhance the rates of fundamental processes such as atom-atom and atom-light scattering when atoms scatter into already occupied states. For non-interacting bosons, the enhancement of light scattering follows directly from the occupation of the {\color{black} atom's} final momentum state. Here, we study scattering between off-resonant light and atoms in a quasi-homogeneous Bose gas with tunable interactions and show that even weak interactions, which do not significantly alter the momentum distribution, strongly affect atom-light scattering. Changes in local atomic correlations suppress the bosonic enhancement under weak repulsive interactions and increase the scattering rate under attractive ones. Moreover, if the interactions are rapidly tuned, light scattering reveals correlation dynamics that are orders of magnitude faster than the collisional dynamics of the momentum{\color{black}-space} populations. Its extreme sensitivity to correlation effects makes off-resonant light scattering a {\color{black} powerful} probe of many-body physics in ultracold atomic gases.
\end{abstract}
\maketitle

Scattering experiments have played a pivotal role in condensed matter physics, from the discoveries of crystal~\cite{Bragg:1913b, Davisson:1927, Brockhouse:1995} and quasi-crystal~\cite{Shechtman:1984} structures to the studies of critical phenomena~\cite{Ford:1965} and superfluidity \cite{Woods:1973}. In a typical experiment, the particles in the probe beam, such as photons, electrons, or neutrons, interact only weakly with the system under investigation, and the correlations in the system are revealed because the interference of different scattering events affects the observed scattering rate. 

Recently, such experiments were used to observe quantum-statistical correlations in ultracold atomic gases~\cite{Deb:2021, Margalit:2021, Sanner:2021, Lu:2023}. If a high-temperature ideal gas is illuminated by an off-resonant laser, the intensity of the scattered light is simply proportional to the number of illuminated atoms. However, at high phase-space density (in quantum-degenerate gases), the innate tendency of identical bosons to bunch and fermions to anti-bunch leads to an enhancement (bosonic stimulation) of scattering for bosons and a suppression (Pauli blocking) for fermions. These effects were predicted more than 30 years ago~\cite{Svistunov:1990-2, Politzer:1991,Javanainen:1994, Morice:1995, Javanainen:1995b, Politzer:1997}, but observed only recently, in harmonically trapped Bose~\cite{Lu:2023} and Fermi~\cite{Deb:2021, Margalit:2021, Sanner:2021} gases. 

Here, we study scattering between off-resonant light and atoms in a quasi-homogeneous Bose gas, and observe an interplay of bosonic statistics and atomic interactions that is not captured by the standard picture of bosonic stimulation. We show that, due to changes in local atomic correlations, even weak and short-ranged atomic interactions can dramatically suppress or further increase the bosonically enhanced scattering rate. These effects are not captured by mean-field theory, which successfully describes the effects of weak interactions for many other observables~\cite{Pitaevskii:2016}. {\color{black} Note that the mean-field wavefunction is, as for an ideal gas, a symmetrized product of single-particle states (which are plane waves for a homogeneous system) and captures purely quantum-statistical correlations, but not any correlations induced by the interactions; to capture the latter, one has to use superpositions of different symmetrized product states.} The sensitivity of light scattering to beyond-mean-field effects makes it a powerful probe of both equilibrium and non-equilibrium many-body physics in quantum gases.


Our experiments are based on a $\potassium$ gas held in an optical box trap~\cite{Gaunt:2013,Navon:2021}, close to the critical temperature for Bose--Einstein condensation, $\Tc$, and with tunable contact interactions characterized by the $s$-wave scattering length $a$. Our trap is formed by intersecting two hollow-tube blue-detuned laser beams with diameters of $35\,\upmu$m  and $100\,\upmu$m (see Fig.~1\textbf{a} and Methods), and we start by preparing $N = (4 - 12) \times 10^5$ atoms in the $\ket{F, m_F} = \ket{1,-1}$ hyperfine ground state. The gas density, $n = (5-15) \, \um^{-3}$, is almost uniform across the cloud, and $\Tc = (140-270)\,$nK. The interactions in our clouds are always weak in the sense that $|a|$ is much smaller than both the typical inter-particle spacing $d=n^{-1/3}$ and the thermal de Broglie wavelength $\lambda=\sqrt{2\pi \hbar^2/(m \kB T)}$, where $\hbar$ is the reduced Planck constant, $m$ the atom mass, and $\kB$ the Boltzmann constant.

We illuminate the gas with an off-resonant laser beam (detuned by $1$\,GHz from the D2 line at \SI{766.7}{\nano \meter}), and detect photons scattered at an angle of \SI{35}{\degree} (Fig.~1\textbf{a}). The laser intensity and pulse duration are chosen such that the cloud is only weakly perturbed (Extended Data Fig.~1), so we probe correlations that are native to the gas and avoid light-induced effects such as super-radiance~\cite{Dicke:1954,Inouye:1999a}. 
In this case, the scattering rate $\Gamma$ at recoil momentum $\hbar \mathbf{Q}$ is given by~\cite{Lewenstein:1994,Ruostekoski:1997b,Javanainen:1999}
\begin{equation}
\begin{split}
   \frac{\Gamma}{\Gamma_0}  = 1  + \frac{1}{N} & \iint  \textrm{d}^3\mathbf{r}_1\, \textrm{d}^3\mathbf{r}_2 ~e^{i\mathbf{Q}\cdot(\mathbf{r}_1-\mathbf{r}_2)} \\ 
 &\times \text{Tr}[\rho~\hat{\psi}^\dagger(\mathbf{r}_1)   \hat{\psi}^\dagger(\mathbf{r}_2)  \hat{\psi}(\mathbf{r}_2) \hat{\psi}(\mathbf{r}_1)]
    \,,
\end{split}
\label{eq:1}
\end{equation}
where $\Gamma_0$ is the single-particle scattering rate, $\hat{\psi}(\mathbf{r})$ the bosonic field operator, and $\rho$ the system's thermal density matrix. Note that here the enhancement factor $E=\Gamma/\Gamma_0$ is simply the structure factor $S(\mathbf{Q})$.

\begin{figure*}   
    \centering
    \includegraphics[width=\textwidth]{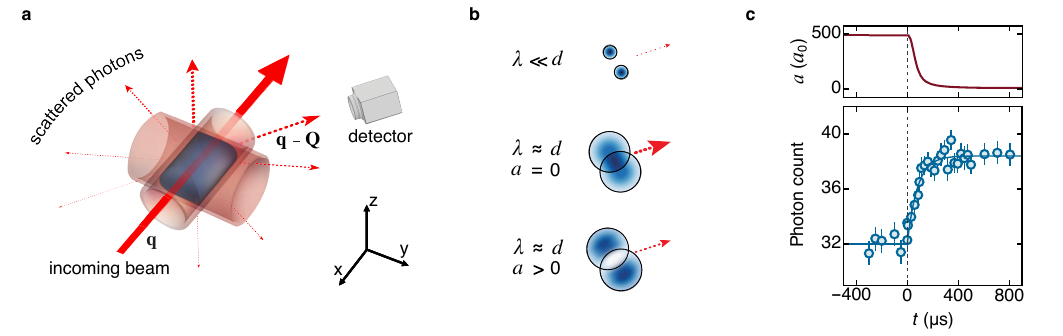}
\caption{ 
\textbf{Bosonic enhancement of atom-light scattering in an interacting gas.} {\textbf a}, 
Experimental concept. We illuminate a quasi-homogeneous Bose gas (blue) held in an optical box trap (red) with an off-resonant laser beam and detect photons scattered at an angle of 35$^{\rm o}$. Scattering of photons from wavevector ${\bf q}$ to ${\bf q} - {\bf Q}$ corresponds to atom recoil $\hbar {\bf Q}$.
\textbf{b},
Cartoon of the effects of quantum degeneracy and atomic interactions on atom-light scattering. 
Here $d$ is the inter-particle spacing (set by the gas density), $\lambda$ is the thermal de Broglie wavelength (the size of the blue atomic wave-packets), and the $s$-wave scattering length $a$ gives the strength of interactions. The scattered-light intensity is indicated by the thickness of the red arrows. In a degenerate non-interacting gas, with $\lambda \gtrsim d$ and $a=0$, scattering is enhanced by interference of light scattered by overlapping wave-packets. However, this overlap and enhancement are reduced by repulsive interactions ($a>0$).
{\bfseries c}, 
Illustration of the local effect of interactions; here $T \approx 1.1\, \Tc$, where $\Tc\approx 200\,$nK is the critical temperature for Bose--Einstein condensation.  Reducing $a$ from $500\,a_0$ to $<50\,a_0$ within $200\,\upmu$s (top panel) enhances atom-light scattering on the same timescale (bottom panel), which is too short to change the global momentum-space occupations or the gas density distribution (see text).
For each measurement, at different times during the interaction ramp, the light pulse was applied for $10\,\upmu$s, and the single-particle scattering rate $\Gamma_0$ corresponds to $27$ photon counts. {\color{black} Each data point represents the average of $107$ independent experimental runs, and the error bars indicate the standard error of the mean (s.e.m.).} 
}.
    \label{fig1}
\end{figure*}

As we sketch in Fig.~1\textbf{b}, in an ideal ($a=0$) gas close to $\Tc$, where $\lambda \approx d$, the scattering rate is enhanced as the overlap of the atomic wave-packets and the bosonic symmetry of the wavefunction lead to enhanced density fluctuations.
In this simple $a=0$ case, and for a uniform gas, the second-order correlator in Eq.~(\ref{eq:1}) factorizes into first-order correlators, and
\begin{equation}
    E = \frac{\Gamma}{\Gamma_0}= \frac{1}{N} \sum_\mathbf{k} N_\mathbf{k} (1 + N_{\mathbf{k} + \mathbf{Q}})\, , 
\label{eq:2}
\end{equation}
where $N_\mathbf{k}$ is the occupation of the momentum state $\mathbf{k}$.

This is the standard textbook result which states that the rate of a physical process is enhanced by a factor of $1+ N_{\rm f}$, where $N_{\rm f}$ is the number of bosons already occupying the final quantum state, and here the relevant states are the final momentum states, $\mathbf{k} + \mathbf{Q}$, of the recoiling atoms. 
The same picture is also commonly used to explain a bosonically enhanced rate of elastic atom-atom scattering, for example during formation of a Bose--Einstein condensate~\cite{Miesner:1998b}.

However, as we also sketch in Fig.~1\textbf{b} (bottom cartoon), repulsive interactions ($a>0$) suppress the wave-packet overlap and the resulting enhancement~\cite{Naraschewski:1999, Lu:2023}. In this case Eq.~(\ref{eq:1}) does not reduce to Eq.~(\ref{eq:2}), and the scattering rate cannot be understood in terms of momentum-space occupations. Physically, the key point is that the momentum states are spatially delocalized, while bosonic enhancement fundamentally arises from spatially local correlations, through the interference of light scattered by two nearby atoms, and only for non-interacting particles the pictures in terms of local correlations and global state occupations are (mathematically) equivalent. 

Indeed, in our experiments the interactions dramatically affect light scattering without significantly changing the momentum-space occupations; in fact, they affect it on a timescale on which $N_\mathbf{k}$ cannot change. We illustrate this in Fig.~1\textbf{c}, by showing how $\Gamma$ grows when the interactions are rapidly turned off. Here $T \approx1.1\,\Tc$ and $\Gamma_0$ corresponds to $27(3)$ photon counts (Methods). Using a magnetic Feshbach resonance at $33.6\,$G, we first prepare the cloud at a relatively large $a=500\,a_0$, where $a_0$ is the Bohr radius. We then sweep the $B$ field to reduce $a$ to $< 50\,a_0$ within $200\,\upmu$s (top panel), and observe that $\Gamma$ grows on the same timescale (bottom panel). For comparison, it takes several milliseconds for the global momentum-space occupations and density distribution to change [the rates of these processes are set, respectively, by the collision rate, $8 \pi n a^2 \sqrt{2\kB T/(\pi m)}$, and the frequency of sound waves, $\propto \sqrt{\kB T/m}$].
This separation of timescales clearly shows that the change in $\Gamma$ arises due to local correlations.

We now turn to a quantitative study of the bosonic enhancement factor $E$ for a range of temperatures, gas densities, and interaction strengths (Fig.~2).

Starting with the case of a quasi-ideal gas ($a=25\,a_0$, corresponding to $a/\lambda < 3 \times 10^{-3}$), in Fig.~2{\bf a} we plot $E$ as a function of $T/\Tc$ for three different densities. Near $\Tc$, in our densest clouds we observe enhancement up to a factor of $2$. 

\begin{figure*}   
    \centering
\includegraphics[width=\textwidth]{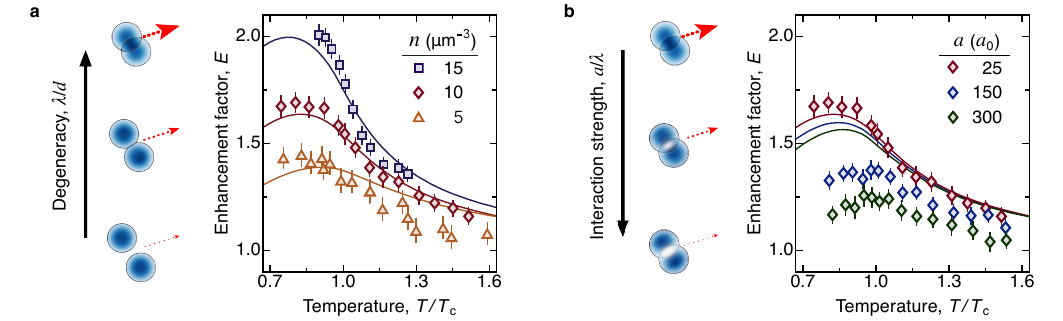}
\caption{
\textbf{Effects of quantum degeneracy and interactions.} {\color{black} Here, the error bars denote the s.e.m.} \textbf{a,} The bosonic enhancement factor, $E=\Gamma/\Gamma_0$, as a function of the reduced temperature $T/\Tc$ for a quasi-ideal gas ($a = 25\,a_0$). The increase of enhancement with the gas density $n$ or when approaching condensation is captured well by mean-field numerical calculations (solid lines). {\color{black} Data points show averages over independent experimental runs, with $132$, $150$ and $100$ repetitions for the lowest to highest density, respectively.} \textbf{b,} $E$ versus $T/\Tc$ for different interaction strengths and fixed $n = \SI{10}{\micro \meter}^{-3}$. Mean-field effects (the differences between the three solid lines) do not explain the dramatic suppression of bosonic enhancement. {\color{black} All data corresponds to the mean of 150 independent experiments. }
} 
\label{fig2}
\end{figure*}

We reproduce these observations in numerical calculations (solid lines in Fig.~2{\bf a}) based on the standard bosonic-stimulation result in Eq.~(\ref{eq:2}). Our simulations take into account that: (i) the gas density depends slightly on the temperature and mean-field interaction energy because the trap walls are not infinitely steep (see Methods and Extended Data Fig.~2), and (ii) Eqs.~(\ref{eq:1},\,\ref{eq:2}) assume spin-preserving scattering, but in our experiments about a quarter of the scattering events are spin-changing (see Methods and Extended Data Fig.~3), and this Raman process is not bosonically enhanced~\cite{Lu:2023}.

However, as we show in Fig.~2{\bf b}, for stronger interactions (at fixed density), bosonic enhancement is dramatically suppressed over our whole temperature range, and only a small fraction of this suppression is explained by the mean-field calculations (solid lines) based on Eq.~(\ref{eq:2}). Note that we normalize $T$ using the experimentally observed $\Tc$; for our range of interaction strengths, $\Tc$ is expected to shift with $a$ by up to a few percent, but this effect is not within our experimental resolution.

We now focus on the interaction effects on $E$, including their dynamics, for fixed $n = 10 \, \um^{-3}$ and $T = \Tc$ (Fig.~3). Here we employ two atomic spin states, $\ket{1,-1}$ and $\ket{1,0}$, which due to different Feshbach resonances~\cite{Roy:2013,Etrych:2023} have different values of $a$ at the same $B$ field (see Fig.~3{\bf a}), and use two-photon (Raman) spin flips to realize sub-\si{\micro\second} interaction quenches (see Methods and Extended Data Fig.~4). This allows us to probe purely beyond-mean-field effects of changes in $a$, on timescales on which the global density and momentum distributions do not change.

In Fig.~3\textbf{b}, we show $E(t)$ for spin flips at three different $B$ fields. In one measurement (blue), $a$ is the same before and after the flip, and we verify that $E$ remains the same.
When we suddenly decrease or increase $a$ (red and green, respectively), the value of $E$ adjusts in $\approx \SI{50}{\micro\second}$, independently of the quench direction; the exponential fits shown by the solid lines give a time constant $\tau = \SI{25(5)}{\micro\second}$.
Since $\tau$ is much longer than the sub-\si{\micro\second} spin flip, and much shorter than the millisecond mean-field timescales, it reflects the intrinsic timescale for the growth or decay of local correlations; note that, in contrast, in Fig.~1\textbf{c} the rate at which $\Gamma$ changes is limited by how fast $a$ changes.

In Fig.~3\textbf{c} we summarize our results for interaction quenches at a range of fields (shaded regions in Fig.~3\textbf{a}) where $a$ in the initial $\ket{1,-1}$ state is always small, $(25 - 40)\,a_0$, 
while $a$ in the final $\ket{1,0}$ state varies between $35\,a_0$ and $750\,a_0$. We measure $E$ at $t=\SI{50}{\micro\second}$ after the spin flip, when the local correlations have (essentially) fully adjusted. Plotting $E$ versus the dimensionless interaction strength $a/\lambda$ shows that a scattering length that is an order of magnitude smaller than the size of the atomic wave packets (and the typical inter-particle separation) is sufficient to almost completely suppress the bosonic enhancement of light scattering.

This suppression is captured well by considering correlation effects at the two-body level~\cite{Naraschewski:1999,Lu:2023}, which effectively amounts to replacing
\begin{equation}
N_{\rm f} \rightarrow N_{\rm f} \left(1 - c \, \frac{a}{\lambda} \right) \, 
\label{eq:3}
\end{equation}
in the standard $1+N_{\rm f}$ expression for the bosonically enhanced scattering events
(see Methods and Extended Data Figs.~5 and 6). In the limit of high temperature and zero recoil momentum, $c= 8\sqrt{2}\approx11.3$, while for our experiments at $\Tc$ we numerically obtain $c\approx 12$, which is coincidentally similar; {\color{black} for $T=\Tc$, our photon-scattering angle of $35^{\rm o}$, and $n$ varying from $5$ to $15 \, \um^{-3}$ (our experimental range), we numerically get $c$ that varies from $\approx 15$ to $\approx 10$}.
A value of $c \gg 1$ reflects the extreme sensitivity of light scattering to the local interaction-induced correlations. In Fig.~3\textbf{c} the solid line shows our calculated steady-state $E$, and the dashed line takes into account the fact that in $2\tau$ the correlations do not yet fully settle. 

\begin{figure*}   
\centering
\includegraphics[width=\textwidth]{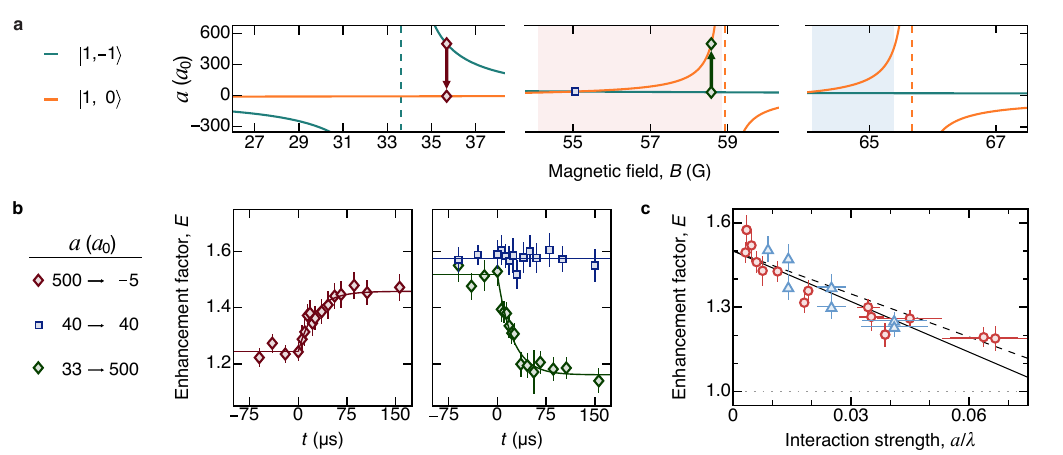}
\caption{
{\bfseries Beyond-mean-field interaction effects.}
Here we fix $n = 10 \, \um^{-3}$ and $T = \Tc$, and perform spin-flips from the $\ket{1,-1}$ to the $\ket{1,0}$ atomic state at different magnetic fields to realize sub-$\upmu$s interaction quenches and study the subsequent evolution of $E$ on microsecond timescales, when only local correlations can change. For all data, {\color{black} the error bars represent the s.e.m. and} $E$ is measured using $4\,\upmu$s-long light-scattering pulses. 
\textbf{a}, $a(B)$ for the two spin states, with the dashed lines indicating Feshbach resonances (where $|a| \rightarrow \infty$). 
The symbols and arrows depict the spin-flips in \textbf{b}, and the shaded regions show the $B$-field ranges, near two different $\ket{1,0}$ resonances, used in \textbf{c}. 
\textbf{b}, Correlation dynamics. Independently of the initial and final value of $a$, the light-scattering rate adjusts to the quench at $t=0$ within the same time of $\approx 50\,\upmu$s; exponential fits (solid lines) give a time-constant $\tau=\SI{25(5)}{\micro\second}$. {\color{black} Here, each data point represents the average of $116$ repetitions.}
\textbf{c}, 
Quasi-steady-state $E$, measured $50\,\upmu$s after the quench from $a<40\,a_0$ in $\ket{1,-1}$ to $a = (35-750)\,a_0$ in $\ket{1,0}$; the symbol colors match the shadings in {\bf a}. 
Plotting $E$ versus $a/\lambda$ shows that a scattering length an order of magnitude smaller than the size of the wave packets and the inter-particle distance (see Fig.~\ref{fig1}{\bf b}) is sufficient to almost completely suppress bosonic enhancement. 
The solid line shows the steady-state $E$ predicted by our beyond-mean-field calculations (see text) and the dashed line includes the correction for the fact that in $2\tau$ the correlations do not yet fully settle. {\color{black} All data is the mean of $123$ independent experimental runs.}
}
\label{fig3}
\end{figure*}

Our dynamical simulations also reproduce the characteristic timescale of $\SI{25}{\micro\second}$ for the change in correlations after quenches from either a small to large $a$ or vice versa (see Methods and Extended Data Fig.~7). Our calculations give that well above $\Tc$ the time for the relaxation of correlations scales as $m\lambda/(\hbar Q)$; this can be understood as thermal dephasing, with the relaxation time corresponding to the time it takes particles with typical thermal velocity $\hbar/(m\lambda)$ to travel the distance $1/Q$, over which the correlations are probed. However, for $T\rightarrow \Tc$ the relaxation time grows (Methods), suggesting sensitivity of light scattering to critical behavior. 

 \begin{figure}   
\centering
\includegraphics[width=\columnwidth]{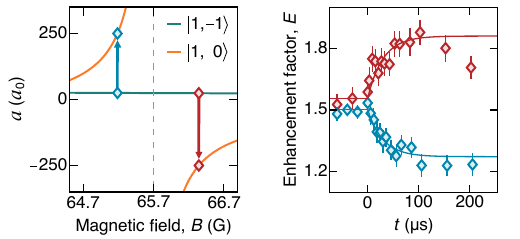}
\caption{
{\bfseries Effect of attractive interactions.}
Left: Experimental protocol. We prepare the gas in a weakly interacting $|1,-1\rangle$ state and then quench at $t=0$ to either $a=250\,a_0$ or $a = - 250\,a_0$ in the $|1,0\rangle$ state. Right: Post-quench dynamics. Attractive interactions increase the bosonic enhancement of atom-light scattering above the ideal-gas level ($E\approx 1.5$). At times of order $\tau \approx 25\,\upmu$s, the interaction effects are essentially symmetric for $a=\pm 250\,a_0$, as theoretically expected to leading order in $a/\lambda$. The solid lines show exponential fits to the data for $t< 110\,\upmu$s with fixed $\tau = 25\,\upmu$s. {\color{black} Each data point is the average of 118 (red) and 87 (blue) independent runs. The error bars show the s.e.m.}
}
\label{fig4}
\end{figure}

Finally, in Fig.~4 we briefly explore the effect of attractive interactions ($a<0$), still for $n = 10 \, \um^{-3}$ and $T = \Tc$, and still starting in the $\ket{1,-1}$ state. Here we use spin flips to $\ket{1,0}$ near the $65.7\,$G resonance to rapidly quench $a$ from $\approx 30\,a_0$ to $\pm 250\,a_0$. We observe that attractive interactions increase the scattering rate above the ideal-gas level ($E \approx 1.5$, see also Fig.~3{\bf c}). Moreover, at post-quench times of order $\tau$, the effects of quenching to $\pm 250\,a_0$ are essentially symmetric, as theoretically expected at leading order in $a/\lambda$; at longer times the $a<0$ dynamics are more complicated and we do not presently understand them. Note that we do not explore stronger attractive interactions because, for the same $|a|$, particle loss due to three-body recombination is generally more severe for $a<0$~\cite{Kraemer:2006,Zaccanti:2009}; this loss is negligible over \SI{200}{\micro\second} at $-250\,a_0$, and at $750\,a_0$, but at $-600\,a_0$ we observe $30\%$ atom loss within \SI{100}{\micro\second}.

In summary, our work sheds new light on the fundamental phenomenon of bosonic stimulation, relevant across many fields, and establishes off-resonant light scattering as a powerful new tool for probing second-order correlations in ultracold atomic gases~\cite{Jeltes:2007, Hung:2011-2, Perrin:2012, Hung:2013, Tenart:2021, Bureik:2025, Brandstetter:2024, deJongh:2024, Xiang:2024, Yao:2024}.
In the future, probing light scattering at different angles would allow studies of correlations on different lengthscales, which could, for example, provide new insights into the critical behavior near $\Tc$. Moreover, our methods are also applicable to far-from-equilibrium systems, such as Bose gases quenched to unitarity (where $a > \lambda, d$)~\cite{Makotyn:2014, Eigen:2018} and turbulent gases~\cite{Henn:2009, Navon:2016}, which are generally less understood; so far these systems have been characterized by their statistically averaged momentum distributions ({\it i.e.}, first-order correlations), while second-order correlations could reveal their fluctuations, including intermittency in quasi-stationary turbulence~\cite{Batchelor:1949, Newell:2001}.

Our work also raises new conceptual questions and invites complementary experiments in other systems.
The suppression and enhancement of light scattering imply a modification of the real-space bunching of bosons, which should be directly observable with quantum-gas microscopes~\cite{Brandstetter:2024, deJongh:2024, Xiang:2024, Yao:2024}.
Interactions should also modify the bosonic enhancement of elastic atom-atom scattering, seen in the formation of a Bose--Einstein condensate~\cite{Miesner:1998b}. Moreover, since in equilibrium the scattering into and out of the condensate are related by detailed balance, this raises the question whether the modification of bosonic stimulation has a thermodynamic signature in the equilibrium condensed fraction.
It would also be interesting to study the effects of contact interactions on the fermionic suppression of light scattering~\cite{Deb:2021, Margalit:2021, Sanner:2021}, and the effects of long-range (dipolar) interactions on both bosonic stimulation and fermionic suppression.
Finally, another intriguing question is whether emission of light that is bosonically stimulated by the occupation of the final photon states (rather than the final atom states) can also be modified by some form of photon-photon~\cite{Firstenberg:2013} or emitter-photon interactions.

{\bf Acknowledgments}\quad We thank Christopher J. Ho, Gevorg Martirosyan, Ji\v{r}\'{i}~Etrych, Martin Gazo, Andrey Karailiev, Yi Jiang, and Martin Zwierlein for discussions and comments on the manuscript. The Cambridge work was supported by EPSRC [Grant No.~EP/Y01510X/1], ERC [UniFlat], and STFC [Grants No.~ST/T006056/1 and No.~ST/Y004469/1]. Z.H. acknowledges support from the Royal Society Wolfson Fellowship. 
The MIT work was supported from the NSF through grant No.~PHY-2208004, from the Center for Ultracold Atoms (an NSF Physics Frontiers Center) through grant No.~PHY-2317134, and the Army Research Office (contract No.~W911NF2410218).  Y.-K.L. is supported by the NTT Research Fellowship.

{\bf Author contributions}\quad K.K., P.H.C.W., and N.D. led the experimental work. Y.Z. led the theoretical work. All authors contributed to the interpretation of the results and writing of the manuscript.

{\bf Competing Interests Statement}\quad The authors declare no competing interests.



%

\newpage

\setcounter{figure}{0} 
\setcounter{equation}{0}

\renewcommand\theequation{S\arabic{equation}} 
\renewcommand\thefigure{\arabic{figure}} 
\renewcommand{\theHfigure}{E\arabic{figure}}
\renewcommand{\figurename}[1]{Extended Data Fig.~}

\clearpage

\section*{METHODS}

\subsection*{Linear response}

In Extended Data Fig.~1 we show the recorded photon count versus the time-integrated pulse intensity, verifying that we work in the linear-response regime.

\begin{figure}[h!]
\centering
\includegraphics[width=\columnwidth]{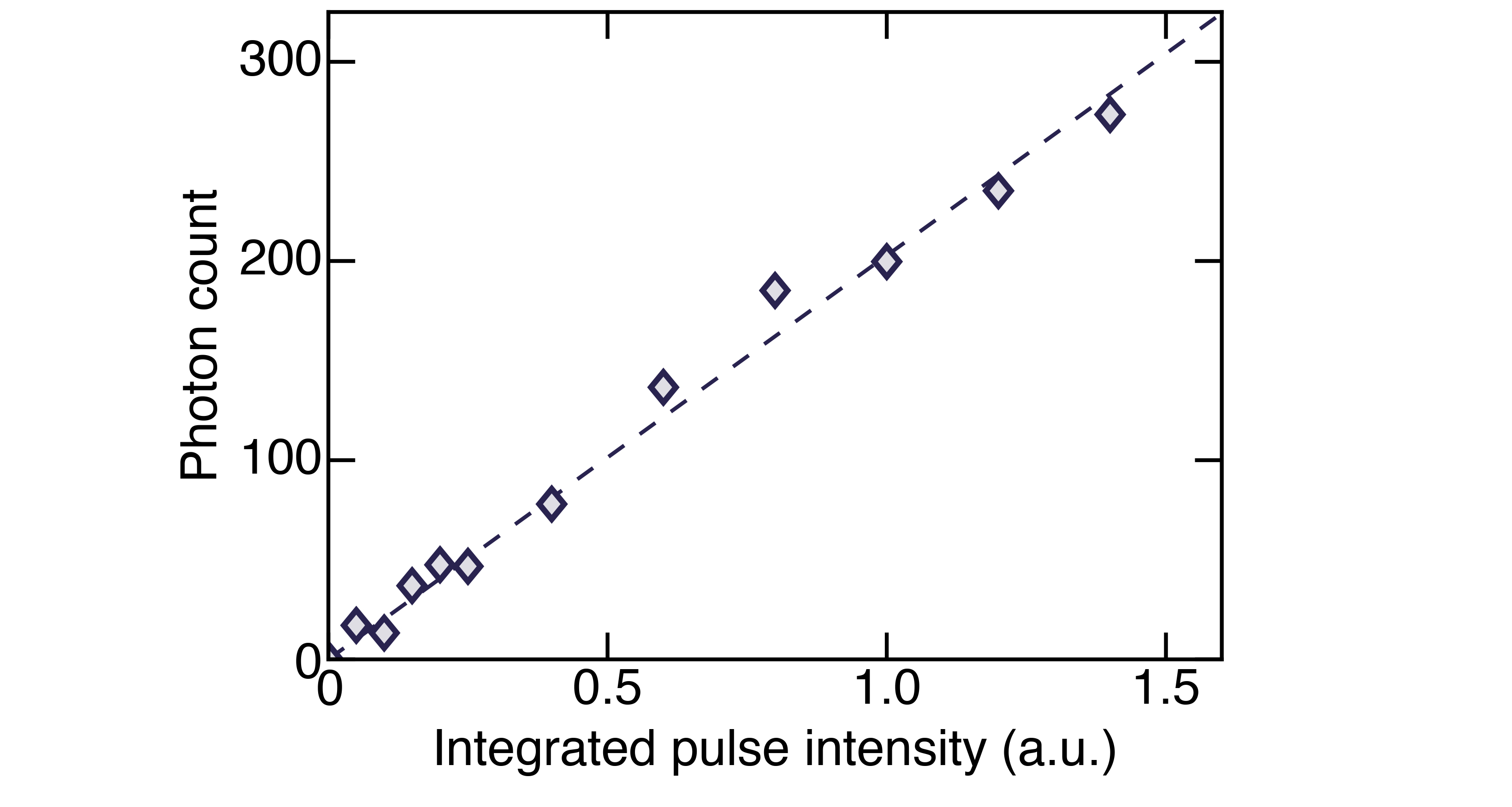}
\caption{\textbf{Linearity of the response.} 
We show the linearity of the recorded photon count with the time-integrated intensity of the atom-light scattering pulse. We always work with light pulses such that the scattered-photon count is $\lesssim 60$. We estimate that this corresponds to $\lesssim 0.1$ photons scattered per atom.
}
\label{fig:si:1}
\end{figure}

\subsection*{Optical box trap}

Our optical box trap is formed by intersecting two hollow-tube $755$\,nm laser beams with diameters of $\approx 35\,\upmu$m and $\approx 100\,\upmu$m, resulting in a trap volume $V\approx 80 \times 10^{3}\,\upmu$m$^3$. To extract the thermodynamic properties of our gas, we take absorption images after \SI{20}{\milli \second} of time-of-flight expansion at weak interactions ($a < 40\,a_0$).

\begin{figure}[h!]   
\centering
\includegraphics[width=\columnwidth]{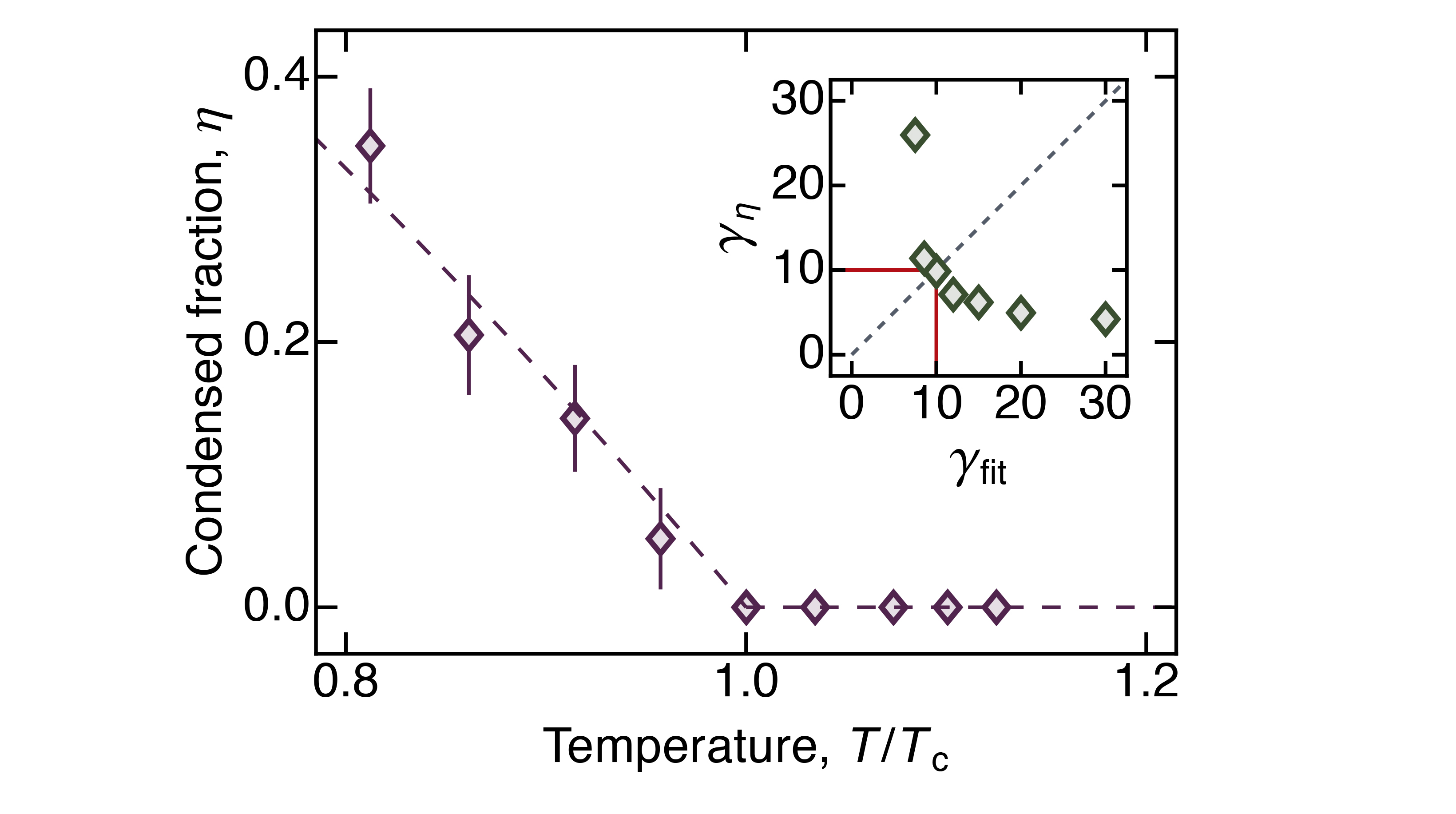}
\caption{
\textbf{Box sharpness.} Main panel: temperature dependence of the condensed fraction, obtained using Eq.~(\ref{eq:si:1}) with $\gamma_{\rm fit} =10$, is fitted well to Eq.~(\ref{eq:si:2}) with the consistent $\gamma_{\eta} =10$ (dashed line). {\color{black} Each data point is the average of 93 independent measurements and the error bars denote the standard deviation.} Inset: $\gamma_{\eta}$ obtained for different $\gamma_{\rm fit}$; the consistency requirement $\gamma_{\eta} = \gamma_{\rm fit}$ (dashed line) gives $\gamma = \gamma_{\eta} = \gamma_{\rm fit} = 10$. 
}
\label{fig:si:2}
\end{figure}

To calibrate the steepness of our trap walls we follow Ref.~\cite{Schmidutz:2014}, assuming an isotropic power-law trapping potential, $U(r) \propto r^\gamma$, for which (assuming an ideal Bose gas) the momentum distribution of the thermal component is
\begin{equation}
f(k)\propto g_{3/\gamma}\left( z \exp\left[-\frac{\hbar^2k^2}{2m \kB T}\right]\right)\, ,
\label{eq:si:1}
\end{equation} 
and the condensed fraction for $T< \Tc$ is
\begin{equation}
\eta = 1 - \left(\frac{T}{\Tc}\right)^{\frac{3}{2} + \frac{3}{\gamma}}\,,
\label{eq:si:2}
\end{equation}
where $g_\alpha(k) = \sum_{x=1}^\infty (k^x/x^\alpha)$ is the polylogarithm function, $z=\exp[{\mu/(\kB T)}]$ the fugacity, and $\mu$ the chemical potential.
We fit $f(k)$ using different values of $\gamma$, denoted $\gamma_{\rm fit}$, and then from the deduced $\eta(T/\Tc)$ extract another value of $\gamma$, denoted $\gamma_{\eta}$. Self-consistency then requires $\gamma_{\eta} = \gamma_{\rm fit}$. As we show in Extended Data Fig.~2, our data is described well by $\gamma = 10$.

\subsection*{Light scattering and collection}

We count the scattered photons using an electron multiplying charge-coupled device (EMCCD) camera (PROHS-1024BX3), with a quantum efficiency of $95\%$, in an imaging system with a numerical aperture of $\approx 0.1$. Hardware binning over $32\times32$ pixels and a gain of $\approx 100$ are used to ensure that photon shot noise dominates over the readout noise. The typical measured photon count corresponds to a single-shot signal-to-noise ratio of about $6$; our measurements are then averaged over at least $50$ repetitions.

For each set of $E$ measurements we normalize $\Gamma$ by the single-particle scattering rate $\Gamma_0$, which we obtain by probing clouds that were allowed to expand in time-of-flight for $\SI{10}{\milli \second}$. Note that the systematic uncertainty in determining $\Gamma$ and $\Gamma_0$ separately is $\lesssim10\%$, but that does not affect their ratio.

\begin{figure}[b]
\centering
\includegraphics[width=\columnwidth]{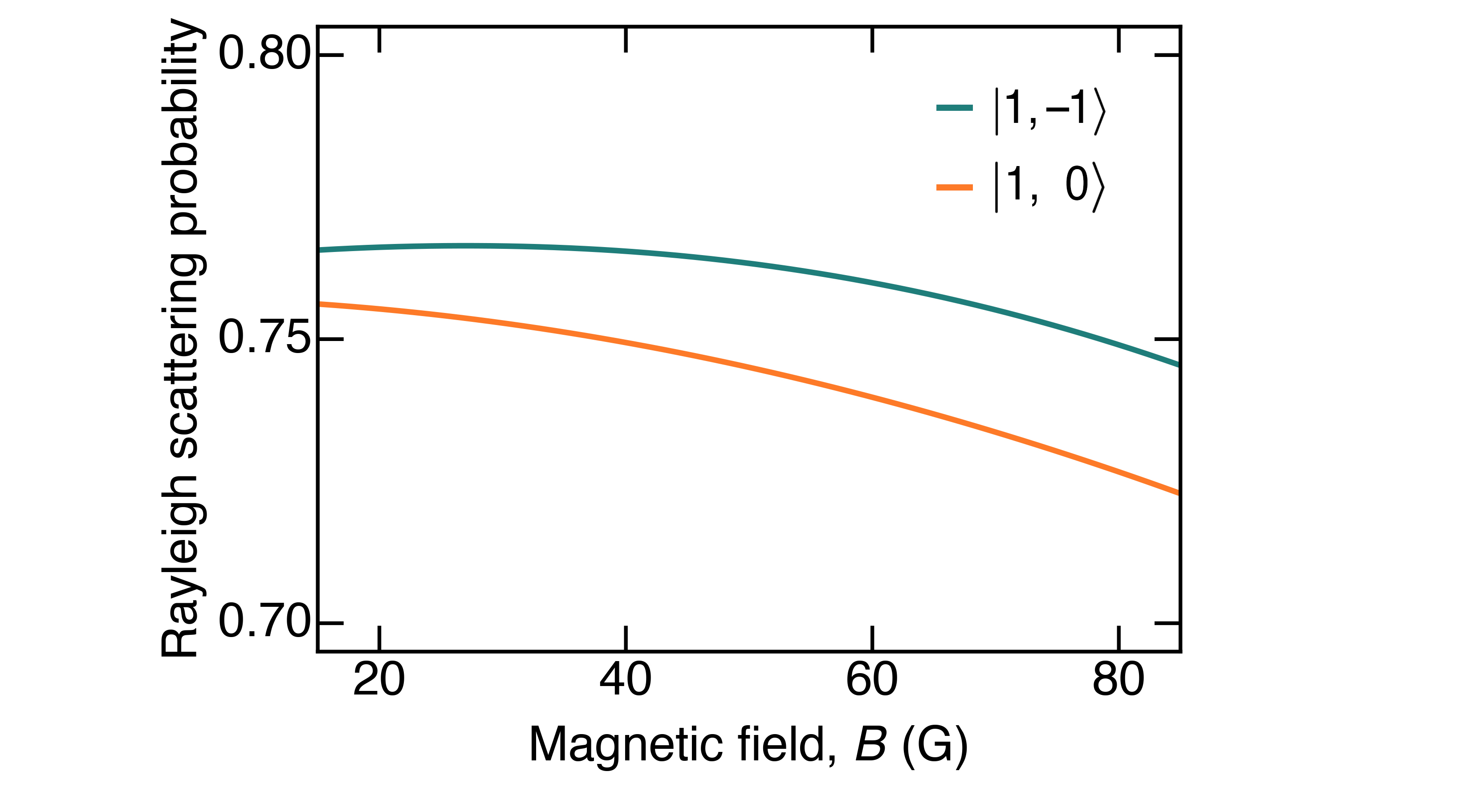}
\caption{\textbf{Probability of Rayleigh scattering.} 
For our experimental parameters, both our spin states, and all $B$ fields used, calculations give that for single-particle scattering ($\lambda \ll d$) the fraction of scattering events that are spin-preserving is always close to $0.75$.}
\label{fig:si:3}
\end{figure}

In our experiments, with the incoming beam detuned $1\,$GHz from the $\SI{766.7}{\nano\metre}$ line and having $\sigma^-$ polarization with respect to its horizontal propagation axis (Fig.~1{\bf a}), the $B$ field oriented along the vertical axis, and the scattered photons collected at an angle of \SI{35}{\degree} in the horizontal plane, both spin-preserving (Rayleigh) and spin-changing (Raman) scattering occur, and only the former is bosonically enhanced~\cite{Lu:2023}. In Extended Data Fig.~3, we show for both our spin states the calculated fraction of scattering events that are spin-preserving in the limit of single-particle scattering.

Finally, note that photon re-absorption (multiple-scattering) is negligible for our system because the mean-free-path of the photons is $> 2\,$cm, much larger than the cloud size.

\subsection{Spin flips}
We use a pair of co-propagating laser beams (detuned by $2$\,GHz from the $\SI{766.7}{\nano\metre}$ line) to implement two-photon Raman transitions between the $\ket{1,-1}$ and $\ket{1,0}$ spin states at different $B$ fields.
In Extended Data Fig.~4, we plot the measured fraction of atoms in each state for a varying Raman-pulse duration, showing a spin-flip ($\pi$-pulse) time of $\approx 0.5\,\upmu$s.

\begin{figure}[h!]
\includegraphics[width=\columnwidth]{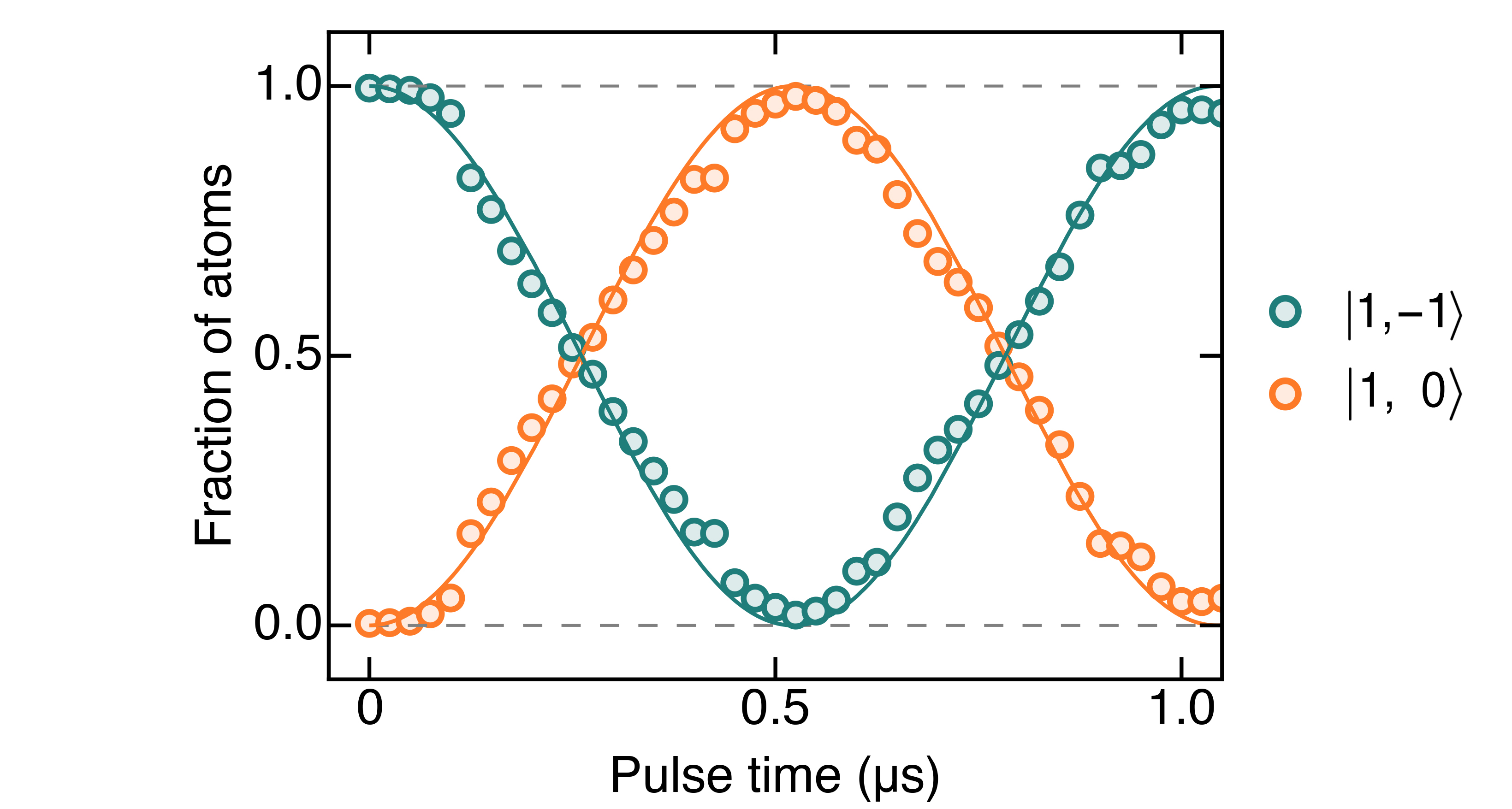}
\caption{\textbf{Spin transfer.} Fraction of atoms in $\ket{1,-1}$ and $\ket{1,0}$ states as a function of the Raman-pulse duration. The solid lines show sinusoidal fits that give a Rabi frequency $\Omega/(2\pi)\approx 1$\,MHz.  
}
\label{fig:si:4}
\end{figure}

\subsection{$S(\mathbf{Q})$ calculations}

Here, we focus on the non-trivial part of the structure factor, $\tilde{S}(\mathbf{Q}) = S(\mathbf{Q})-1$:
\begin{equation}
\begin{split}
  \tilde{S}(\mathbf{Q}) &=\,    \frac{1}{N}  \iint   ~e^{i\mathbf{Q}\cdot(\mathbf{r}_1-\mathbf{r}_2)} \\ 
 &\times \text{Tr}\,\left[\rho~\hat{\psi}^\dagger(\mathbf{r}_1)   \hat{\psi}^\dagger(\mathbf{r}_2)  \hat{\psi}(\mathbf{r}_2) \hat{\psi}(\mathbf{r}_1)\right] \textrm{d}^3\mathbf{r}_1\, \textrm{d}^3\mathbf{r}_2
    \,.
\end{split}
\label{eq:si:3}
\end{equation}

At the two-body level, $s$-wave interaction with scattering length $a$ modifies the non-interacting two-body eigenstates
\begin{equation}
|\mathbf{k}_1,\mathbf{k}_2\rangle_0 = e^{i \mathbf{k}_1 \cdot \mathbf{r}_1} e^{i \mathbf{k}_2 \cdot \mathbf{r}_2},
\label{eq:si:4}
\end{equation}
with single-particle momenta $\hbar\mathbf{k}_1$ and $\hbar\mathbf{k}_2$, to
\begin{equation}
    |\mathbf{k}_1,\mathbf{k}_2\rangle = e^{i\mathbf{K} \cdot\mathbf{R}} \left[ e^{i\mathbf{k} \cdot \mathbf{r}} - a \frac{e^{i k r}}{r}\right],
    \label{eq:si:5}
\end{equation}
where $\mathbf{R} = (\mathbf{r}_1 + \mathbf{r}_2) / 2$ and $\mathbf{K} = \mathbf{k}_1 + \mathbf{k}_2$ are the center of mass variables, and $\mathbf{r} = \mathbf{r}_1 - \mathbf{r}_2$ and $\mathbf{k} = (\mathbf{k}_1 - \mathbf{k}_2)/2$ are the relative variables.

The interacting and non-interacting states are related by the Moller operator $\Omega_+$~\cite{Taylor:2006}:
\begin{equation}
  |\mathbf{k}_1,\mathbf{k}_2\rangle = \Omega_+ |\mathbf{k}_1,\mathbf{k}_2\rangle_0 \, .
  \label{eq:si:6}
\end{equation}
For bosons, in terms of field operators:
\begin{equation}
\begin{split}
    \Omega_+ = \mathbb{I} \, + \, \frac{i}{2}\iiiint
    &
    \alpha(\mathbf{k}_1, \mathbf{k}_2\xrightarrow{}\mathbf{k}_1', \mathbf{k}_2') \,
    \\
    &\times \hat{\psi}^\dagger(\mathbf{k}_1') \hat{\psi}^\dagger(\mathbf{k}_2') 
    \hat{\psi}(\mathbf{k}_2) 
    \hat{\psi}(\mathbf{k}_1)
    \\
    & \times
    \frac{\textrm{d}^3 \mathbf{k}_1}{(2\pi)^3}
    \frac{\textrm{d}^3 \mathbf{k}_2}{(2\pi)^3}
    \frac{\textrm{d}^3 \mathbf{k}_1'}{(2\pi)^3}
    \frac{\textrm{d}^3 \mathbf{k}_2'}{(2\pi)^3}
    \,,
    \label{eq:si:7}
\end{split}
\end{equation}
where $\mathbb{I}$ is the identity operator and 
\begin{equation}
\begin{split}
    \alpha(\mathbf{k}_1, \mathbf{k}_2 \xrightarrow{} \mathbf{k}_1', \mathbf{k}_2') = \, &(2\pi)^3 \, \delta(\mathbf{k}_1'+ \mathbf{k}_2' - \mathbf{k}_1 - \mathbf{k}_2) \\
    & \times 4\pi i  \frac{a}{k'^2 - k^2 - i\epsilon}\,,
    \label{eq:si:8}
\end{split}
\end{equation}
where $k'=|\mathbf{k}_1'- \mathbf{k}_2'|/2$ and $k=|\mathbf{k}_1 -\mathbf{k}_2|/2$.

For a dilute gas, we consider only pairwise corrections to the many-body wave function, which is the leading-order approximation. This amounts to expressing the thermal density matrix $\rho$ as~\cite{Naraschewski:1999}
\begin{equation}
    \rho \simeq \Omega_+ \, \rho_0 \, \Omega_+ ^\dagger \,,
    \label{eq:si:9}
\end{equation}
where $\rho_0$ is the non-interacting thermal density matrix.

Using Eqs.~(\ref{eq:si:7})-(\ref{eq:si:9}) and denoting $\tilde{S}_0(\mathbf{Q})$ as the $a=0$ value of $\tilde{S}(\mathbf{Q})$,
we obtain:
\begin{equation}
\begin{split}
&\tilde{S} (\mathbf{Q}) =  \\  &\tilde{S}_0(\mathbf{Q)} 
                    - \frac{16 \pi a}{n Q^2} \iint 
                    \frac{N_{\mathbf{k}_1} N_{\mathbf{k}_2}}{1+{\mathbf{Q}\cdot(\mathbf{k}_1 - \mathbf{k}_2)/Q^2}}
                    \frac{{\textrm d}^3\mathbf{k}_1}{(2\pi)^3} \frac{{\textrm d}^3\mathbf{k}_2}{(2\pi)^3}
                    \,,
    \label{eq:si:10}
\end{split}
\end{equation}
\begin{figure} [t!]  
\centering
\includegraphics[width=\columnwidth]{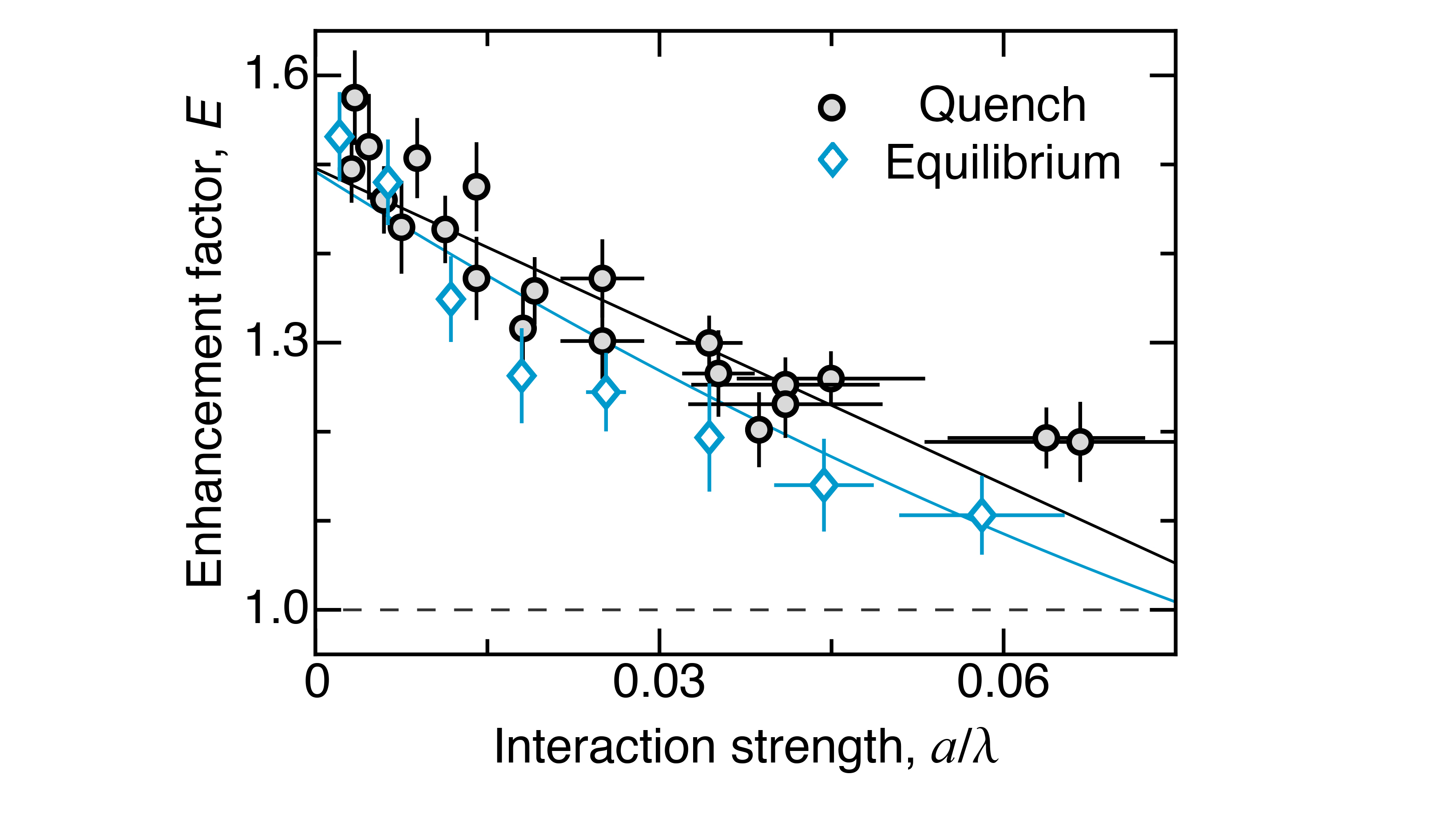}
\caption{\textbf{Additional equilibrium measurements.}
For $n=10\,\um^{-3}$ and $T=\Tc$, we plot equilibrium $E(a/\lambda)$ (blue), obtained as in Fig.~\ref{fig2}, alongside the post-quench data from Fig.~\ref{fig3}{\bf c} (black). The solid lines show our calculations for the two experimental scenarios (see text). {\color{black} Each data point is the mean of 123 (quench) and 100 (equilibrium) independent runs and the error bars show the s.e.m.}
}
\label{fig:si:5}
\end{figure}

where $N_{\bf k}= 1/(z^{-1}\exp[\hbar^2k^2/(2m \kB T)]-1)$. Note that this approach is formally equivalent to first-order perturbation theory using the contact interaction potential.


Also note that in the limit of high temperature ($\lambda/d\ll 1$) and low recoil momentum ($Q\lambda \ll 1$) Eq.~(\ref{eq:si:10}) reduces to the result previously obtained in~\cite{Lu:2023}:
\begin{equation}
\tilde{S}(\mathbf{Q} \rightarrow 0) = \tilde{S}_0(\mathbf{Q}\rightarrow 0)\left(1 - 8\sqrt{2} \, a/\lambda\right) \, .
    \label{eq:si:11}
\end{equation}

$\tilde{S}(0)$ quantifies {\it global} density fluctuations (in the presence of a reservoir) and is proportional to isothermal compressibility, which can also be calculated from the {\it mean-field} energy of an interacting gas~\cite{Dalibard:2022}. Interestingly, this means that one can infer long-wavelength fluctuations from the mean-field energy, even though they are not captured by the mean-field wavefunction. However, this thermodynamic connection is not generalizable for $\tilde{S}(Q)$ at arbitrary $Q$, which measures {\it local} fluctuations on lengthscales $1/Q$.

To compare our ($Q>0$) calculations to our measurements of $E$, we: (i)~account for our beam polarisation, and (ii)~use the mean-field (MF) density distribution within the local-density approximation to calculate the local scattering rate, and then integrate it over the trap.

Specifically, for Fig.~3\textbf{c}, to calculate the scattering rate at the final $a$ after the quench, we use the MF density distribution for the initial small $a$ (in state $\ket{1,-1}$), because on the timescale of the measurements the density distribution does not adjust. In contrast, in equilibrium, the density distribution at the final $a$ would also be slightly different. In Extended Data Fig.~5 we show such equilibrium measurements (taken as in Fig.~2), for the same $n = 10 \, \um^{-3}$ and $T=\Tc$. 
In this case, the total reduction of $E$ with $a/\lambda$ is slightly larger.

Finally, in Extended Data Fig.~6 we show our equilibrium calculations for a range of $a/\lambda$ and $T/\Tc$ values, for our setup and $n = 10 \, \um^{-3}$.


\subsection{Quench dynamics}

In Extended Data Fig.~7 we compare our data from Fig.~3{\bfseries b} to numerical calculations of the evolution of the structure factor after an interaction quench, and also show additional simulations for different $Q\lambda$ and $T/\Tc.$ In these calculations we assume a homogeneous cloud and quenches to or from $a=0$.

For the $a\to 0$ case, starting with the interacting thermal density matrix [Eq.~(\ref{eq:si:9})], the dynamics are simply given by the evolution under the non-interacting Hamiltonian $H_0$:
\begin{equation}
    \rho(t) = e^{-i H_0 t/\hbar} \left( \Omega_+ \, \rho_0 \, \Omega_+^\dagger \right)\, e^{i H_0 t/\hbar} \, .
    \label{eq:si:12}
\end{equation}

\begin{figure}[t!] 
\centering
\includegraphics[width=\columnwidth]{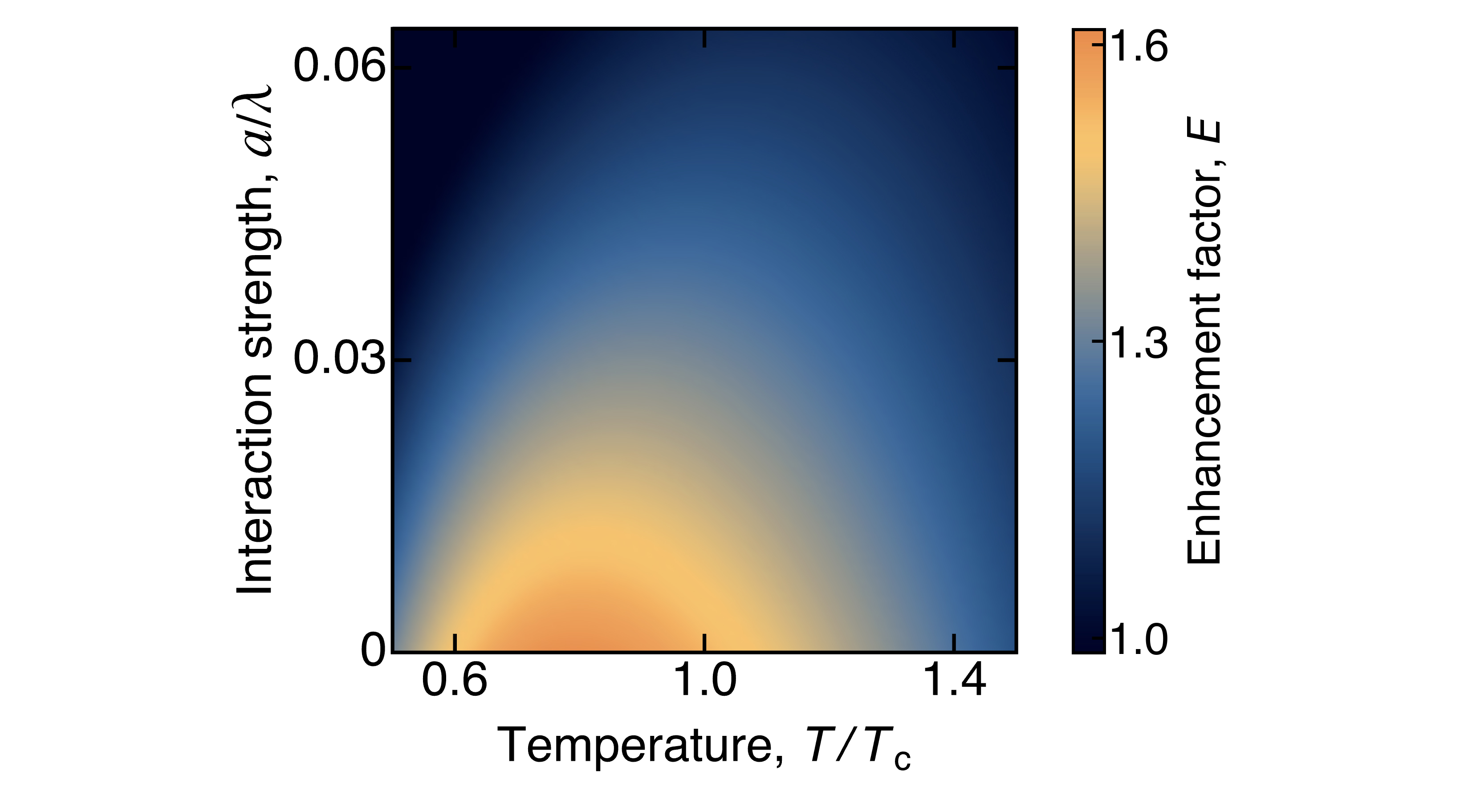}
\caption{
\textbf{Additional equilibrium calculations.} We plot calculated $E$ versus $T/\Tc$ and $a/\lambda$ for $n= 10 \, \um^{-3}$.
}
\label{fig:si:6}
\end{figure}

Dropping the $Q$ labels and writing
\begin{equation}
     \tilde{S}(t) = \Delta (t) + \tilde{S}_0 \, ,
     \label{eq:si:13}
\end{equation}
such that $\Delta (0) = \tilde{S} - \tilde{S}_0$ (which is negative for $a>0$) and $\Delta (\infty) = 0$, we get:
\begin{equation}
\begin{split}
\Delta(t&) =
- \frac{16 \pi a}{n Q^2} \iint 
\frac{N_{\mathbf{k}_1} N_{\mathbf{k}_2}}
{1+{\mathbf{Q}\cdot(\mathbf{k}_1 - \mathbf{k}_2)/Q^2}}
\\
&\times \cos\left(\frac{\hbar Q^2 \left[1+\mathbf{Q}\cdot (\mathbf{k}_1 - \mathbf{k}_2)/Q^2\right] t}{m}\right) \frac{{\rm d}^3\mathbf{k}_1}{(2\pi)^3} \frac{{\rm d}^3\mathbf{k}_2}{(2\pi)^3}   \,.
\end{split}
\label{eq:si:14}
\end{equation}

For the $0\to a$ case, we start with $\rho_0$ and propagate it with the interacting Hamiltonian. Treating the interactions using first-order perturbation theory gives
\begin{equation}
     \tilde{S}(t) =  \tilde{S} - \Delta (t) \, ,
     \label{eq:si:15}
\end{equation}
again with the $\Delta (t)$ given in Eq.~(\ref{eq:si:14}), {\it i.e.}, at this level of approximation the dynamics are completely symmetric.

It is instructive to rewrite Eq.~(\ref{eq:si:14}) as
\begin{equation}
\Delta(t) = - \frac{16\pi \hbar a n}{m} \int_{t}^{\infty} \sin\left( \frac{\hbar Q^2t'}{m}\right)\, \left\vert g^{(1)}\left(\frac{\hbar Qt'}{m}\right)\right\vert ^2 {\rm d} t'\, ,
\label{eq:si:16}
\end{equation}
where 
\begin{equation}
g^{(1)}(r) = \frac{1}{N}\int \text{Tr}\left[ \rho \;\hat{\psi}^\dagger(\bm{r}'+\bm{r})\;\hat{\psi}(\bm{r}')\right] {\rm d}^3\mathbf{r'} 
\label{eq:si:17}
\end{equation}
is the normalized first-order correlation function, which decays over the correlation length $\xi$. This shows that the temporal decay of $\Delta(t)$ is directly related to the spatial decay of $g^{(1)}(r)$, with the characteristic relaxation time set by $m \xi/(\hbar Q)$. Well above $\Tc$, where $g^{(1)}(r) = \exp(-\pi r^2/\lambda^2)$, so $\xi$ is of order $\lambda$, the characteristic relaxation time $\tau_0=m\lambda/(\hbar Q)$ can be interpreted as the time it takes for particles with typical thermal velocity $\hbar/(m\lambda)$ to travel across the lengthscale $1/Q$, over which the correlations are probed by the light scattering. However, as $T \rightarrow \Tc$, the relaxation dynamics slow down due to the growth of $\xi$~\cite{Kardar:2007}.

\begin{figure*}[t!]
\centering
\includegraphics[width=\textwidth]{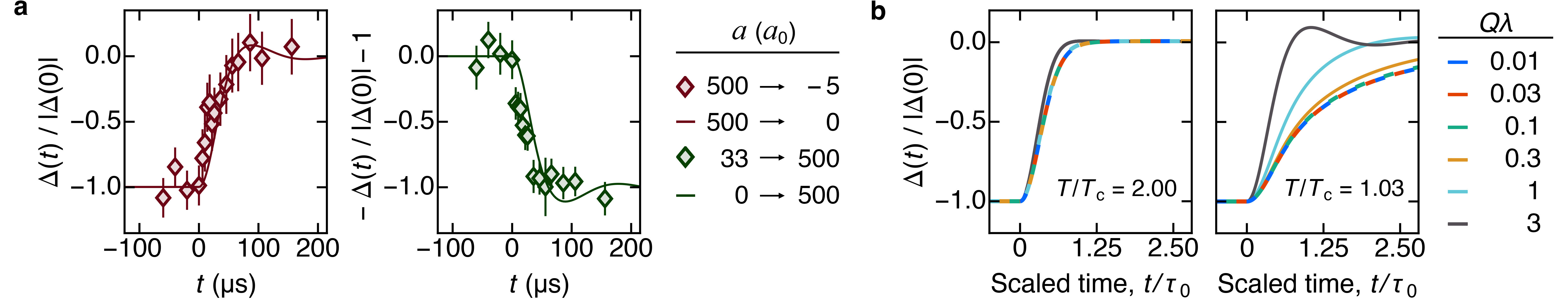}
\caption{\textbf{Quench dynamics.} \textbf{a, } Our calculations (solid lines) of $ \Delta(t)/ |\Delta(0)|$ capture the data from Fig.~\ref{fig3}\textbf{b} (diamonds). Note that the calculations and the data are each normalized with their respective $|\Delta(0)|$. {\color{black}Here, each data point represents the average of $116$ repetitions and the error bars are the s.e.m.}
\textbf{b,} 
Well above $\Tc$ (left), the correlation length $\xi$ is of order $\lambda$ and $\Delta(t)/|\Delta(0)|$ is for $Q\lesssim 1/\lambda$ a universal function of $t/\tau_0$, where $\tau_0=m\lambda/(\hbar Q)$. Close to $\Tc$ (right), $\xi$ grows, the dynamics are universal only for $Q \lesssim 1/\xi$ (here $Q\lesssim 0.1/\lambda$), and these universal dynamics are slower.
}
\label{fig:si:7}
\end{figure*}

Also note that for $Q\ll1/\xi$, the sine function in Eq.~(\ref{eq:si:16}) can be linearized, and the relaxation dynamics for different $Q$ follow a universal trajectory
\begin{equation}
    \Delta(t)
    =  - 16\pi\, n\lambda^3\, \frac{a}{\lambda} \, \int_t^\infty \frac{t'}{\tau_0} \left|g^{(1)}\left(\lambda \frac{t'}{\tau_0}\right)\right|^2~\frac{\text{d} t'}{\tau_0} \, .
    \label{eq:si:18}
\end{equation}
For $T/\Tc\gg1$, this gives $\Delta (t) = \Delta(0) \exp(-2\pi (t/\tau_0)^2)$, with the universal regime extending to $Q\simeq 1/\lambda$.
Approaching $\Tc$, the universal function develops a heavier tail and the extent of the universal regime shrinks (see Extended Data Fig.~7{\bf b}).
For $Q\gtrsim 1/\xi$, the dynamics are more intricate, with $\Delta(t)$ exhibiting damped coherent oscillations. Such oscillatory behavior is predicted for our measurements at $\Tc$, but experimentally we do not resolve these small oscillations within our errors. 

{\bf Data availability}\quad The data that support the findings of this study are available in the Apollo repository (\url{https://doi.org/10.17863/CAM.118185}).  Any additional information is available  from the corresponding authors upon reasonable request.

\twocolumngrid

\onecolumngrid

\end{document}